\documentclass[aps, prl, twocolumn, superscriptaddress, final]{revtex4}

\usepackage[utf8]{inputenc}
\usepackage[T1]{fontenc}
\usepackage{amsmath}
\usepackage{amssymb}
\usepackage{float}
\usepackage{graphicx}

\usepackage{upgreek}

\usepackage{color}
\usepackage{hyperref}
\hypersetup{colorlinks, linkcolor=blue}

\begin{document}

\title{Two-color pulse compounds in waveguides with a zero-nonlinearity point}

\author{O. Melchert}
\email{melchert@iqo.uni-hannover.de}
\affiliation{Leibniz Universit\"at Hannover, Institute of Quantum Optics (IQO), Welfengarten 1, 30167 Hannover, Germany}
\affiliation{Cluster of Excellence PhoenixD (Photonics, Optics, and Engineering - Innovation Across Disciplines), Welfengarten 1A, 30167 Hannover, Germany}

\author{S. Bose}
\affiliation{Cluster of Excellence PhoenixD (Photonics, Optics, and Engineering - Innovation Across Disciplines), Welfengarten 1A, 30167 Hannover, Germany}
\affiliation{Leibniz Universit\"at Hannover, Institue of Photonics (IOP), Nienburger Str. 17, 30167 Hannover}

\author{S. Willms}
\affiliation{Leibniz Universit\"at Hannover, Institute of Quantum Optics (IQO), Welfengarten 1, 30167 Hannover, Germany}
\affiliation{Cluster of Excellence PhoenixD (Photonics, Optics, and Engineering - Innovation Across Disciplines), Welfengarten 1A, 30167 Hannover, Germany}

\author{I. Babushkin}
\affiliation{Leibniz Universit\"at Hannover, Institute of Quantum Optics (IQO), Welfengarten 1, 30167 Hannover, Germany}
\affiliation{Cluster of Excellence PhoenixD (Photonics, Optics, and Engineering - Innovation Across Disciplines), Welfengarten 1A, 30167 Hannover, Germany}

\author{U. Morgner}
\affiliation{Leibniz Universit\"at Hannover, Institute of Quantum Optics (IQO), Welfengarten 1, 30167 Hannover, Germany}
\affiliation{Cluster of Excellence PhoenixD (Photonics, Optics, and Engineering - Innovation Across Disciplines), Welfengarten 1A, 30167 Hannover, Germany}

\author{A. Demircan}
\affiliation{Leibniz Universit\"at Hannover, Institute of Quantum Optics (IQO), Welfengarten 1, 30167 Hannover, Germany}
\affiliation{Cluster of Excellence PhoenixD (Photonics, Optics, and Engineering - Innovation Across Disciplines), Welfengarten 1A, 30167 Hannover, Germany}

\date{\today}

\begin{abstract}
We study incoherently coupled two-frequency pulse compounds in waveguides with single zero-dispersion and zero-nonlinearity points. 
In such waveguides, supported by a negative nonlinearity, soliton dynamics can be obtained even in domains of normal dispersion.
We demonstrate trapping of weak pulses by solitary-wave wells, 
forming nonlinear-photonics meta-atoms,
and molecule-like bound-states of pulses.
We study the impact of Raman effect on these pulse compounds, finding that, depending on the precise subpulse configuration, they decelerate, accelerate, or are completely unaffected.
Our results extend the range of systems in which two-frequency pulse compounds can be expected to exist and demonstrate further unique and unexpected behavior.
\end{abstract}

\maketitle

\paragraph{Introduction}
The incoherent interaction of optical pulses is a central concern in nonlinear optics. 
%
%
For instance, strong and efficient control of light pulses has been shown for a soliton, which induces a strong refractive index barrier that cannot be surpassed by quasi group-velocity matched waves located in a domain of normal dispersion \cite{Demircan:PRL:2013,Demircan:OL:2014}, resulting in mutual repulsion. 
This mechanism is naturally supported by the supercontinuum generation process \cite{Driben:OE:2010,Demircan:APB:2014}.
%
%
%
A transfer of this concept to waveguides supporting group-velocity matched copropagation of pulses in separate domains of anomalous dispersion yields an entirely attractive interaction \cite{Melchert:PRL:2019}.
In this case, cross-phase modulation (XPM) induced potential wells provide a binding mechanism that enable molecule-like bound states of pulses.
They form a single compound pulse, consisting of two subpulses at vastly different frequencies.
%
These objects were previously studied by putting emphasis on their frequency-domain representation, 
showing that a soliton can act as localized trapping potential with discrete level spectrum \cite{Melchert:PRL:2019}, supporting the formation of two-frequency pulse compounds in cases where both subpulse-amplitudes are of similar size \cite{Melchert:PRL:2019,Melchert:OL:2021}. Perturbations of various type where studied in this context \cite{Melchert:SR:2021,Willms:PRA:2022,Oreshnikov:arXiv:2022}.
%
A complementing approach in terms of a multi-scales analysis, putting emphasis on the representation in the time domain, 
showed that they form a class of generalized dispersion Kerr solitons which can be described using the concept of a meta-envelope \cite{Tam:OL:2019}.
Such two-color solitons were recently verified experimentally in mode-locked laser cavities \cite{Lourdesamy:NP:2021,Mao:NC:2021}.
%
%
%
%
Here, we extend the range of systems in which such pulse compounds can be observed. We consider waveguides with a single zero-dispersion point and a single zero-nonlinearity point, where the nonlinear coefficient is negative in the domain of normal dispersion.
This setup allows for group-velocity matching within a large range of frequencies, and allows insight into the complex interplay of sign changing nonlinear and dispersive effects.
%
Photonic-crystal fibers with frequency dependent nonlinearity with the above properties can be obtained by doping with nanoparticles \cite{Driben:OE:2009,Bose:PRA:2016,Bose:JOSAB:2016,Arteaga:PRA:2018,Linale:OL:2020,Hernandez:WRCM:2022}.
Noble gas filled hollow-core waveguides also offer the possibility to have a negative refractive index within a domain of normal dispersion \cite{Junnarkar:OC:2000}.
For a model system with the above properties, we
demonstrate the existence of trapped states in solitary-wave wells, 
show that two-frequency pulse compounds with mutually bound subpulses of similar amplitudes are supported, 
and discuss the dynamics of such pulse complexes in presence of the Raman effect.
The latter leads to the surprising finding that, when the center frequency of the solitary wave-well shifts, a trapped state of higher order can transit into the ground-state.
For our analysis we consider two-frequency pulse compounds for which the subpulses can be well distinguished in the frequency domain, so that their mutual interaction can be described by an incoherent interaction stemming from XPM alone.

\begin{figure}[t]
\centering{\includegraphics[width=\linewidth]{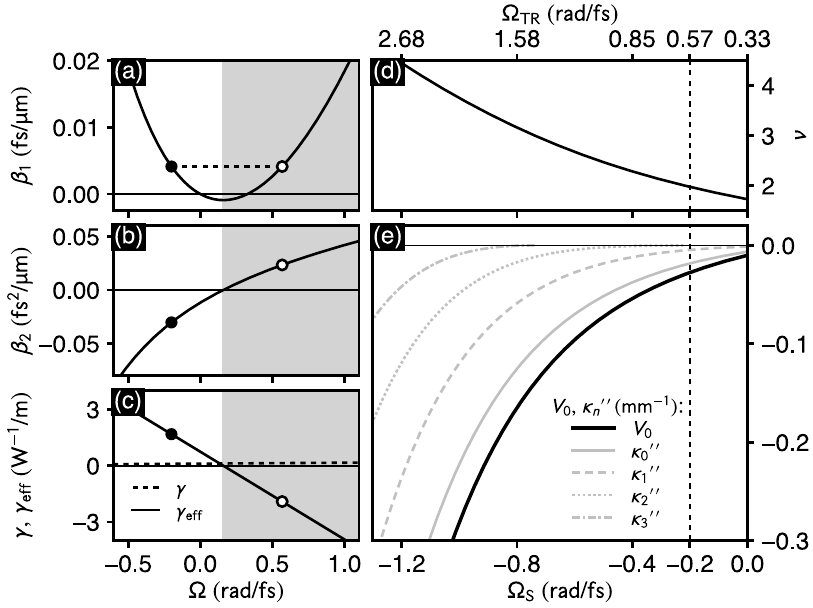}}
\caption{
Specifics of the model.
(a) Group-delay, (b) group-velocity dispersion, and, (c) effective nonlinear coefficient. 
Dot and circle indicate a pair of group-velocity matched pulses. 
Domain of normal dispersion is shaded gray.
(d) Potential strength as function of soliton center frequency $\Omega_{\rm{S}}$. 
Labels on top indicate $\Omega_{\rm{TR}}$, i.e.\ group-velocity matched frequencies at which trapped states exist.
(e) Wavenumber eigenvalues $\kappa^{\prime\prime}_n$ and potential depth $V_0=\min(V)$ as function of $\Omega_{\rm{S}}$. Vertical dashed line indicates the 
pair of frequencies in (a-c).
\label{fig:FIG01}}
\end{figure}

\paragraph{Generalized nonlinear Schrödinger equation.}
Subsequently, we model pulse propagation in waveguides with frequency-dependent nonlinearity in terms of the generalized nonlinear Schrödinger equation (GNSE) \cite{Agrawal:BOOK:2019,Bose:PRA:2016,Zhao:PRA:2022}
\begin{align}
i\partial_z A = &-\sum_{n\geq2} \frac{\beta_n}{n!} (i\partial_t)^n A - (1-f_R)\gamma_{\mathrm{eff}} |A|^2 A \notag \\ 
&-  f_R \gamma A \int_0^{\infty} h_R(t^\prime) |A(z,t-t^\prime)|^2~{\mathrm{d}}t^\prime,
\label{eq:GNLS}
\end{align}
for a complex-valued envelope $A=A(z,t)$. Therein, time $t$ is measured in a reference frame moving with the group velocity at $\omega_0\approx 2.2559\,\mathrm{rad/fs}$, and $z$ is the propagation distance.
Following Ref.~\cite{Zhao:PRA:2022},
the dispersion coefficients are taken as
$\beta_2 = -1.183\times 10^{-2}\,\mathrm{fs^2/\upmu m}$, 
$\beta_3 = 8.10383\times 10^{-2}\,\mathrm{fs^3/\upmu m}$,
$\beta_4 = -9.5205\times 10^{-2}\,\mathrm{fs^4/\upmu m}$,
$\beta_5 = 0.20737\,\mathrm{fs^5/\upmu m}$, 
$\beta_6 = -0.53943\,\mathrm{fs^6/\upmu m}$,
$\beta_7 = 1.3486\,\mathrm{fs^7/\upmu m}$,
$\beta_8 = -2.5495\,\mathrm{fs^8/\upmu m}$, 
$\beta_9 = 3.0524\,\mathrm{fs^9/\upmu m}$, and
$\beta_{10} = -1.7140\,\mathrm{fs^{10}/\upmu m}$.
As function of the angular frequency detuning $\Omega=\omega-\omega_0$,
they define the propagation constant $\beta(\Omega)=\sum_{n=2}^{10} \beta_n \Omega^n/n!$, with relative group delay $\beta_1(\Omega)=\partial_\Omega \beta(\Omega)$ [Fig.~\ref{fig:FIG01}(a)] and group-velocity dispersion $\beta_2(\Omega)=\partial_\Omega^2\beta(\Omega)$ [Fig.~\ref{fig:FIG01}(b)].
%
The nonlinear coefficients are modeled as 
$\gamma(\Omega) = \gamma_0 + \gamma_1 \Omega$, with $\gamma_0 = 0.11\,\mathrm{W^{-1}/m}$ and $\gamma_1=4.8728\times 10^{-5}\,\mathrm{ps\,W^{-1}/m}$, and as
$\gamma_{\rm{eff}}(\Omega) = \gamma_{0,\mathrm{eff}} + \gamma_{1,\mathrm{eff}} \Omega$, with $\gamma_{0,\mathrm{eff}} = 0.7453\,\mathrm{W^{-1}/m}$, and $\gamma_{1,\mathrm{eff}}=-4.6822\times 10^{-3}\,\mathrm{ps \,W^{-1}/m}$ [Fig.~\ref{fig:FIG01}(c)].
For the considered parameters, the zero-disperion point,
defined by $\beta_2(\Omega_{\rm{ZDP}})=0$, and the 
zero-nonlinearity point, defined by $\gamma_{\rm{eff}}(\Omega_{\rm{ZNP}})=0$, are at $\Omega_{\rm{ZDP}}\approx\Omega_{\rm{ZNP}}\approx 0.16\,\mathrm{rad/fs}$.
The Raman effect is included as
$h_{R}(t) = (\tau_1^2 + \tau_2^2)\tau_1^{-1} \tau_2^{-2}\,\exp(-t/\tau_2)\,\sin(t/\tau_1)$
with $f_R=0.18$, $\tau_1=12.2\,\mathrm{fs}$, and $\tau_2=32\,\mathrm{fs}$ \cite{Blow:JQE:1989}. 
%
%
%
%
For the solution of Eq.~(\ref{eq:GNLS}) with $f_R=0.18$ we use a split-step Fourier method \cite{Agrawal:BOOK:2019}. When neglecting the Raman effect, i.e.\ for $f_R=0$, we use the conservation quantity error method \cite{Heidt:JLT:2009,Melchert:CPC:2022}.
To assess time-frequency interrelations within $A(z,t)$, we use the spectrogram
$P_{S}(t,\Omega) = \left|\int A(z,t^\prime) 
\exp\left[-(t^\prime-t)^2/2\sigma^2 - i \Omega t^\prime\right]~{\rm d}t^\prime\right|^2$ \cite{Melchert:SFX:2019},
employing a Gaussian window function with root-mean-square width $\sigma=50\,\mathrm{fs}$.

\paragraph{Coupled nonlinear Schrödinger equations.}
In search of incoherently coupled two-frequency pulse compounds, we intentionally neglect the Raman effect and
consider complex-valued envelopes $A_1$, and $A_2$, of two group-velocity (GV)
matched pulses with a vast frequency gap [Fig.~\ref{fig:FIG01}(a)],
described by the two coupled nonlinear
Schrödinger equations (NSEs) 
\begin{subequations}\label{eq:CNSE}
\begin{align}
&i \partial_z \,A_1 - \frac{\beta_2^\prime}{2} \partial_t^2 \,A_1 + \gamma^\prime \left( |A_1|^2 + 2 |A_2|^2 \right) A_1=0, \label{eq:CNSE1} \\ 
&i \partial_z \,A_2 - \frac{\beta_2^{\prime\prime}}{2} \partial_t^2 \,A_2 + \gamma^{\prime\prime} \left( |A_2|^2 + 2 |A_1|^2 \right) A_2=0.
\label{eq:CNSE2}
\end{align}
\end{subequations}
The incoherently coupled NSEs (\ref{eq:CNSE}) further neglect higher orders of dispersion as well
as four-wave mixing contributions between the two pulses. 
Their mutual interaction is included via XPM alone.
Considering the pair of GV matched frequencies $\Omega_1=-0.20\,\mathrm{rad/fs}$, and
$\Omega_2=0.57\,\mathrm{rad/fs}$ [Fig.~\ref{fig:FIG01}(a)], 
yields
$\beta_2^{\prime} = -0.0303\,\mathrm{fs^2/\upmu m}$, $\gamma^{\prime}=1.68\,\mathrm{W^{-1}/m}$, 
$\beta_2^{\prime\prime}=0.0234\,\mathrm{fs^2/\upmu m}$, and $\gamma^{\prime\prime}= -1.91\,\mathrm{W^{-1}/m}$.
This distinguishes the present setup from earlier ones where $\beta_2^{\prime},\,\beta_2^{\prime\prime}<0$, and $\gamma^{\prime},\,\gamma^{\prime\prime}>0$ \cite{Melchert:PRL:2019}.
%
%
Below we look for solutions to Eqs.~(\ref{eq:CNSE}) in the form 
\begin{align}
A_1(z,t) = U_1(t) e^{i \kappa^\prime z}, \quad \text{and} \quad A_2(z,t) = U_2(t) e^{i \kappa^{\prime\prime} z}, \label{eq:ansatz}
\end{align}
wherein $U_1$, $U_2$ are real-valued envelopes, and $\kappa^\prime$,
$\kappa^{\prime\prime}$ are the corresponding wave numbers.
Substituting Eqs.~(\ref{eq:ansatz}) into Eqs.~(\ref{eq:CNSE}) yields
the two coupled ordinary differential equations (ODEs)
\begin{subequations}\label{eq:ODEs}
\begin{align}
&\ddot U_1 - \frac{2}{\beta_2^{\prime}}\left[  \gamma^\prime \left( |U_1|^2 + 2 |U_2|^2\right) - \kappa^{\prime}\right] U_1 = 0, \label{eq:2ODE1} \\ 
&\ddot U_2 - \frac{2}{\beta_2^{\prime\prime}}\left[  \gamma^{\prime\prime} \left( |U_2|^2 + 2 |U_1|^2\right) - \kappa^{\prime\prime}\right] U_2 = 0,\label{eq:2ODE2}
\end{align}
\end{subequations}
where the dots denote derivatives with respect to time.

\begin{figure}[t]
\centering{\includegraphics[width=\linewidth]{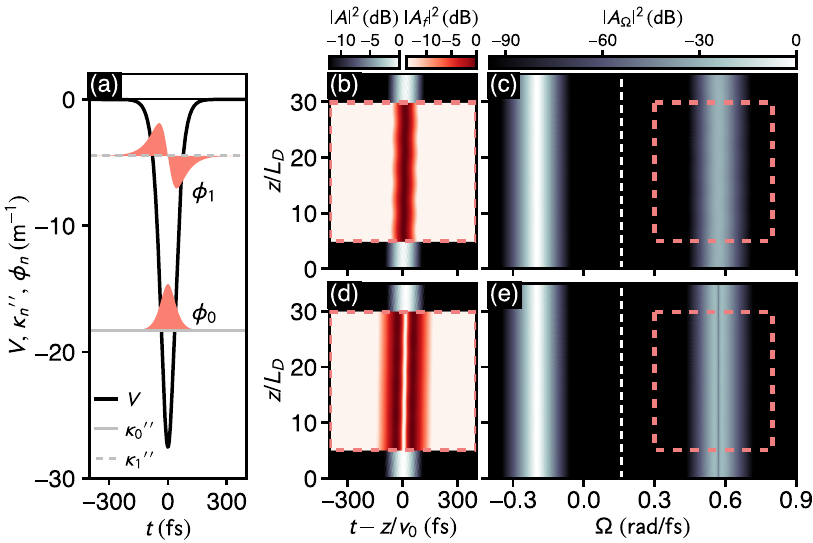}}
\caption{Solitary-wave well with two trapped states.
(a) Trapping potential $V$, wavenumber eigenvalues $\kappa^{\prime\prime}_n$, and eigenfunctions $\phi_n$, $n=0,1$.
(b) Time-domain propagation dynamics of the soliton and its trapped state $n=0$.
(c) Corresponding spectrum.
Filtered view in (b) details the time-domain view of the frequency range enclosed by the dashed box in (c).
(d,e) Same as (b,c) for $n=1$.
\label{fig:FIG02}}
\end{figure}

\paragraph{Trapped states.}
Imposing the condition $\max(U_2)\ll \max(U_1)$ decouples Eqs.~(\ref{eq:ODEs}): assuming Eq.~(\ref{eq:2ODE1}) to describe a freely propagating soliton $U_1(t) = \sqrt{P_0} \,{\mathrm{sech}}(t/t_0)$ with $P_0=|\beta_2^{\prime}|(\gamma^{\prime}\,t_0^2)^{-1}$ and $\kappa^{\prime}=\gamma^{\prime} P_{0}/2$, Eq.~(\ref{eq:2ODE2}) takes the form of a Schrödinger-type eigenvalue problem 
\begin{align}
 - (\beta_2^{\prime\prime}/2)\ddot{\phi}_n + V(t) \phi_n =  \kappa_n^{\prime\prime}~\phi_n,\label{eq:SEV}
\end{align}
with trapping potential $V(t)=2 \gamma^{\prime\prime} P_0\, {\mathrm{sech}}^2(t/t_0)$.
Since $\beta_2^{\prime\prime}>0$ at $\Omega_2=0.57\,\mathrm{rad/fs}$, the attractive nature of $V$ is enabled by $\gamma^{\prime\prime}<0$ [Fig.~\ref{fig:FIG01}(b,c)]. 
%
%
In Eq.~(\ref{eq:SEV}), the wavenumber eigenvalues are real-valued and satisfy $\kappa_n^{\prime\prime}<0$. 
To each eigenvalue corresponds an eigenfunction $\phi_n(t)$ with $n$ zeros.
%
In analogy to the P\"oschl-Teller potential in one-dimensional quantum scattering theory \cite{Landau:BOOK:1981,Lekner:AJP:2007}, which can be solved exactly,
we write 
$V(t)=-\nu\,(\nu+1)\,\beta_2^{\prime\prime}(2 t_0^{2})^{-1}\,{\mathrm{sech}}^2(t/t_0)$ with strength-parameter
$\nu=-1/2 + \left[1/4 + 4| (\gamma^{\prime\prime}/\gamma^{\prime})(\beta_2^{\prime}/\beta_2^{\prime\prime}
)| \right]^{1/2}$. 
The number of trapped states is $N_{\rm{TR}}=\lfloor \nu \rfloor + 1$, where $\lfloor \nu \rfloor$ is the integer part of $\nu$, and the wavenumber eigenvalues are $\kappa_n^{\prime\prime}=-\beta_2^{\prime\prime} \,(2t_0^2)^{-1}\,(\nu-n)^2$, with $n=0,\ldots,\lfloor \nu \rfloor$.
%
Equation~(\ref{eq:SEV}) suggests an analogy to quantum mechanics, with $\phi_n$ assuming
the role of the wavefunction of a fictitious
particle of mass $m=1/\beta_2^{\prime\prime}$, confined to a localized trapping potential.
The quantized number of trapped states is akin to an atomic number,
and a bare soliton, with none of its trapped states occupied, resembles the nucleus of an one-dimensional atom.
By this analogy, the soliton along with its trapped states represents a nonlinear-photonics meta-atom.
The variation of the potential-strength $\nu$ and the discrete level spectrum of the solitary-wave well $V$ as function of the soliton center frequency $\Omega_{\rm{S}}$ are shown in Figs.~\ref{fig:FIG01}(d,e): for decreasing $\Omega_{\mathrm{S}}$, the trapping potential induced by the soliton features an increasing number of bound states.
An example for the choice $\Omega_{\rm{S}}=-0.20\,\mathrm{rad/fs}$ and $t_0=50\,\mathrm{fs}$, with $\Omega_{\rm{TR}}=0.57\,\mathrm{rad/fs}$ and $\nu \approx 1.98$ [Fig.~\ref{fig:FIG01}(d)] is detailed in Fig.~\ref{fig:FIG02}. There exist $N_{\rm{TR}}=2$ trapped states at $(\kappa_0^{\prime\prime},\,\kappa_1^{\prime\prime})=(-18.29, -4.47)\,\mathrm{m^{-1}}$, given by 
$\phi_0(t)\propto {\mathrm{sech}}^\nu(t/t_0)$, and $\phi_1(t)\propto {\mathrm{sech}}^{\nu-1}(t/t_0)\,{\mathrm{tanh}}(t/t_0)$ [Fig.~\ref{fig:FIG02}(a)].
In the vicinity of $\Omega_{\mathrm{TR}}$,
due to $\kappa_0^{\prime\prime},\,\kappa_1^{\prime\prime}<0$, a wavenumber-gap separates the trapped states from linear waves with propagation constant $\beta^{\prime\prime} = (\beta_2^{\prime\prime}/2) (\Omega-\Omega_{\mathrm{TR}})^2 \geq 0$.
%
%
The stable propagation of intitial conditions $A_0(t)=U_1(t) e^{-i \Omega_{\rm{S}}t} + \phi_n(t) e^{-i \Omega_{\rm{TR}}t}$, with weak trapped states of amplitude
 $\max(|\phi_n|)=0.05 \sqrt{P_0}$, $n=0,1$,
in terms of Eq.~(\ref{eq:GNLS}) in absence of the Raman effect ($f_R=0$) is demonstrated in Figs.~\ref{fig:FIG02}(b-e). 
To account for the change in group-velocity of the soliton in presence of a linear variation of $\gamma$ \cite{Haus:OL:2001}, we consider $v_0^{-1}=\beta_1(\Omega_{\mathrm{S}}) + \gamma_{1,\rm{eff}} P_0$.

\begin{figure}[t]
\centering{\includegraphics[width=\linewidth]{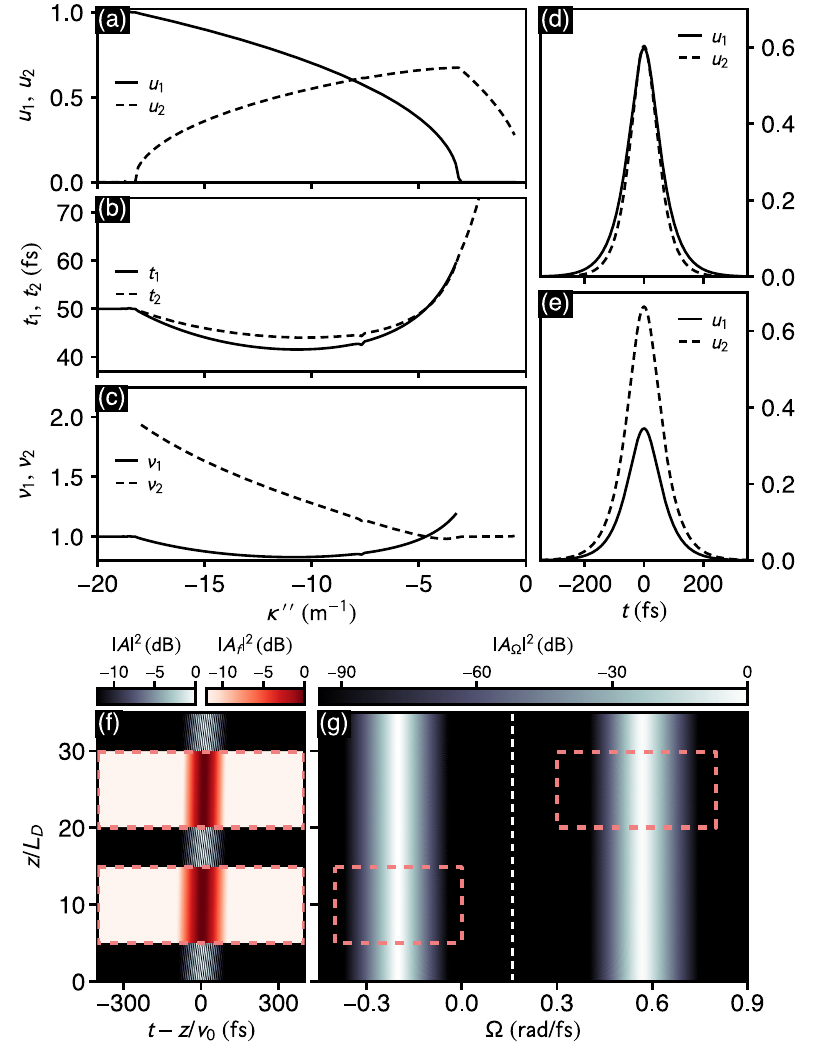}}
\caption{Incoherently coupled two-color pulse compounds.
(a-c) Paramterized solution of Eqs.~(\ref{eq:ODEs}) (see text for parameters). 
(a) Scaled amplitudes $u_n = U_{0,n}/\sqrt{P_0}$, (b) pulse duration $t_n$, and, (c) shape exponent $\nu_n$, $n=1,2$.
(d) Pulse pair for $\kappa^{\prime\prime}=-7.99\,\mathrm{m^{-1}}$, and, (e) pulse pair for $\kappa^{\prime\prime}=-4.68\,\mathrm{m^{-1}}$.
(f) Time-domain propagation dynamics of the pulse pair in (d).
(g) Corresponding spectrum.
Filtered views in (f) detail the time-domain view of the frequency ranges enclosed by the dashed boxes in (g).
\label{fig:FIG03}}
\end{figure}

%
\paragraph{Simultaneous solution of the coupled ODEs.}
%
Solitary-wave solutions of the coupled nonlinear Eqs.~(\ref{eq:ODEs}) beyond the above linear limit yield two-frequency pulse compounds of Eq.~(\ref{eq:GNLS}). 
Under suitable conditions, such solutions can be specified analytically \cite{Haelterman:OL:1993,Silberberg:OL:1995,Afanasjev:OL:1989,Pelinovsky:PRE:2000,Melchert:OL:2021}. 
%
%
However, in order to obtain solutions for general parameter settings, Eqs.~(\ref{eq:ODEs}) need to be solved numerically. This is, e.g., possible via 
shooting methods \cite{Haelterman:PRE:1994,Mitchell:PRL:1997},
spectral renormalization methods \cite{Ablowitz:OL:2005,Lakoba:JCP:2007}, conjugate gradient methods \cite{Lakoba:PD:2009,Yang:JCP:2009}, or Newton methods \cite{Dror:JO:2016}.  
We here employ a Newton method employing a boundary value Runge-Kutta algorithm \cite{Kierzenka:ACM:2001}. 
So as to systematically study solutions to Eqs.~(\ref{eq:ODEs}) 
we set $\kappa^{\prime}=|\beta_2^{\prime}|(2 t_0^2)^{-1}$ with $t_0=50\,\mathrm{fs}$, and start at the location $\kappa^{\prime\prime}=-20\,\mathrm{m^{-1}}$ in parameter space, i.e.\ below the lowest eigenvalue obtained from Eq.~(\ref{eq:SEV}). In this case we expect $U_2$ to vanish, and $U_1$ to yield a fundamental soliton $U_1(t)=\sqrt{P_0}\,{\mathrm{sech}}(t/t_0)$ with $P_0=|\beta_2^{\prime}|(\gamma^{\prime}\,t_0^2)^{-1}$.
We set initial trial functions with parity similar to the soliton and the lowest lying trapped state, and continue the obtained solutions to larger values of $\kappa^{\prime\prime}$. 
The resulting solutions are of the form $U_n(t)=U_{0,n}\,{\mathrm{sech}}^{\nu_n}(t/t_n)$, $n=1,2$,
with parameters summarized in Figs.~\ref{fig:FIG03}(a-c).
%
Consistent with our results above, we find that a weak nonzero solution $U_2$ with $t_2=t_0$ and $\nu_2 \approx 1.98$ originates at $\kappa^{\prime\prime} \approx -18.3\,\mathrm{m^{-1}}$.
Let us point out that the above choice of $\max(\phi_n)/\sqrt{P_0}=0.05$ indeed characterises weak trapped states [Fig.~\ref{fig:FIG03}(a)]. 
For $\kappa^{\prime\prime}>-18.3\,\mathrm{m^{-1}}$, the amplitude of $U_1$ continuously decreases while that for $U_2$ increases. 
Above $\kappa^{\prime\prime}\approx -4\,\mathrm{m^{-1}}$, $U_1$ vanishes and $U_2$ describes a fundamental soliton with wavenumber $\kappa^{\prime\prime}$. 
Let us note that at $\kappa^{\prime\prime}\approx -4.68\,\mathrm{m^{-1}}$ we find a pair of solutions with hyperbolic-secant shape $U_n=U_{0,n}{\mathrm{sech}}(t/t_0)$, $n=1,2$ [Fig.~\ref{fig:FIG03}(e)], i.e.\ a two-color soliton pair as in Ref.~\cite{Melchert:OL:2021}.
%
%
The stable propagation of an initial condition 
$A_0(t)=U_1(t)e^{-i\Omega_{\rm{S}}t}+U_2(t)e^{-i\Omega_{\rm{TR}}t}$ with $U_{0,1} \approx U_{0,2}$ [$\kappa^{\prime\prime}=-8\,\mathrm{m^{-1}}$; Fig.~\ref{fig:FIG03}(d)] in terms of Eq.~(\ref{eq:GNLS}) with $f_R=0$ is demonstrated in Figs.~\ref{fig:FIG03}(f,g).
To account for the change in group-velocity of the pulse compound in Fig.~\ref{fig:FIG03}(f),
we consider $v_0^{-1}=\beta_1(\Omega_{\mathrm{S}}) + \gamma_{1,\rm{eff}} (U_{0,1}^2 + 2 U_{0,2}^2)$, extending the group-velocity correction of Ref.~\cite{Haus:OL:2001} to two-color pulse compounds.

\begin{figure}[t]
\centering{\includegraphics[width=\linewidth]{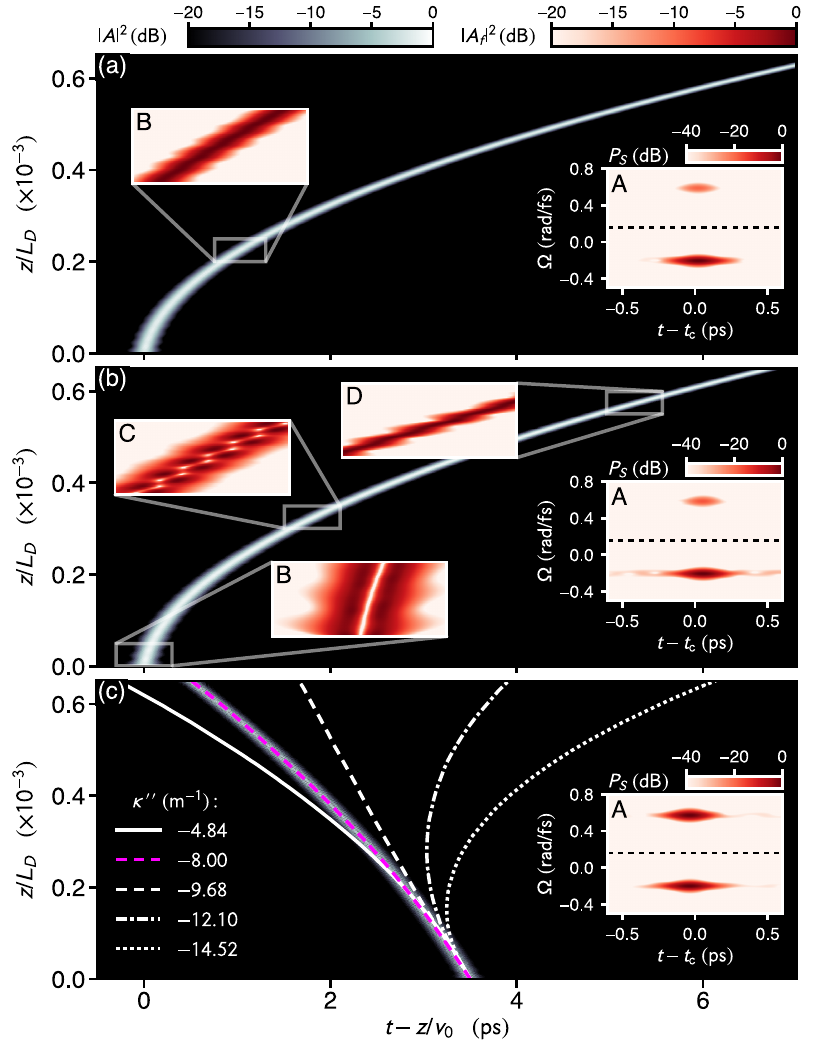}}
\caption{Perturbation by the Raman effect.
(a) Propagation dynamics of a soliton and a weak trapped state of order $n=0$  ($L_D=t_0^2/|\beta^{\prime}_2|$).
(b) Same for $n=1$.
(c) Propagation dynamics of the pulse compound of Fig.~\ref{fig:FIG03}(d).
Inset labeled A shows a spectrogram at $z/L_D = 600$, with $t_c$ indicating the peak-location of the pulse compound. Further insets are detailed in the text.
\label{fig:FIG04}}
\end{figure}

\paragraph{Perturbation by the Raman effect.}
%
We next assess the impact of the Raman effect on the propagation dynamics of the above pulse compounds.
In Fig.~\ref{fig:FIG04}(a) we show a fundamental soliton and a weak trapped state of order $n=0$, propagating under Eq.~(\ref{eq:GNLS}). While the soliton experiences a self-frequency-shift, resulting in a deceleration in the time-domain, the trapped state remains bound by the trapping potential, see the spectrogram in Fig.~\ref{fig:FIG04}(a) (inset A).
Let us note that the level-spectrum of the solitary-wave well is affected by the soliton's frequency downshift [Fig.~\ref{fig:FIG01}(e)]. 
While the soliton decelerates, the trapped state starts to oscillate within the trapping potential (inset B). 
This deceleration induced oscillation within the solitary-wave well bears an unexpected consequence when considering the trapped state of order $n=1$ [Fig.~\ref{fig:FIG04}(b)]:
upon propagation, the initially swift oscillations (inset B) grow in size (inset C) until finally, the trapped state transitions into a trapped state of order $n=0$ (inset D).  The shape-conversion of the mode from $n=1\rightarrow 0$ is also evident in the spectrogram in Fig.~\ref{fig:FIG04}(b) (inset A).
During this transition, a small amount of radiation emanates from the localized pulses.
When considering instances of incoherently coupled two-color pulse compounds [Figs.~\ref{fig:FIG03}(a-c)], the Raman effect can have different consequences [Fig.~\ref{fig:FIG04}(c)]: when $U_{0,1}>U_{0,2}$, the pulse compound decelerates ($\kappa^{\prime\prime} = -14.52\,\mathrm{m^{-1}}$); 
when $U_{0,1} < U_{0,2}$, the pulse compound accelerates ($\kappa^{\prime\prime} = -4.84\,\mathrm{m^{-1}}$); in an intermediate parameter range where $U_{0,1}\approx U_{0,2}$, the pulse compound is nearly unaffected ($\kappa^{\prime\prime} = -9.68\,\mathrm{m^{-1}}$). The latter is a result of the deceleration of one subpulse being counterbalanced by an acceleration of its binding partner.

\paragraph{Summary and conclusions.}
In conclusion, we have demonstrated the existence of two-color pulse compounds in waveguides with a single zero-dispersion point and adequate frequency-dependent nonlinearity.
A strong mutual binding of two group-velocity matched pulses at vastly different center frequencies, earlier demonstrated for waveguides with two domains of anomalous dispersion \cite{Melchert:PRL:2019}, can, in the present case, be achieved by having $\gamma<0$ in a domain where $\beta_2>0$.
The reported study extends the range of systems that support two-color pulse compounds, and allows to understand the complex propagation dynamics reported in a recent study on higher-order soliton evolution in a photonic crystal fiber with one zero-dispersion point and frequency dependent nonlinearity \cite{Zhao:PRA:2022}. Instances of such two-color pulse compounds can readily be identified in the propagation studies reported in Ref.~\cite{Zhao:PRA:2022}.

\subsection*{Acknowledgements}

The authors acknowledge financial support from 
Deutsche Forschungsgemeinschaft within the Cluster of Excellence PhoenixD
(EXC 2122, projectID 390833453).




\bibliography{references}

\begin{thebibliography}{40}
\expandafter\ifx\csname natexlab\endcsname\relax\def\natexlab#1{#1}\fi
\expandafter\ifx\csname bibnamefont\endcsname\relax
  \def\bibnamefont#1{#1}\fi
\expandafter\ifx\csname bibfnamefont\endcsname\relax
  \def\bibfnamefont#1{#1}\fi
\expandafter\ifx\csname citenamefont\endcsname\relax
  \def\citenamefont#1{#1}\fi
\expandafter\ifx\csname url\endcsname\relax
  \def\url#1{\texttt{#1}}\fi
\expandafter\ifx\csname urlprefix\endcsname\relax\def\urlprefix{URL }\fi
\providecommand{\bibinfo}[2]{#2}
\providecommand{\eprint}[2][]{\url{#2}}

\bibitem[{\citenamefont{Demircan et~al.}(2013)\citenamefont{Demircan,
  Amiranashvili, Br\'ee, and Steinmeyer}}]{Demircan:PRL:2013}
\bibinfo{author}{\bibfnamefont{A.}~\bibnamefont{Demircan}},
  \bibinfo{author}{\bibfnamefont{S.}~\bibnamefont{Amiranashvili}},
  \bibinfo{author}{\bibfnamefont{C.}~\bibnamefont{Br\'ee}}, \bibnamefont{and}
  \bibinfo{author}{\bibfnamefont{G.}~\bibnamefont{Steinmeyer}},
  \bibinfo{journal}{Phys. Rev. Lett.} \textbf{\bibinfo{volume}{110}},
  \bibinfo{pages}{233901} (\bibinfo{year}{2013}).

\bibitem[{\citenamefont{Demircan
  et~al.}(2014{\natexlab{a}})\citenamefont{Demircan, Amiranashvili, Br\'ee,
  Morgner, and Steinmeyer}}]{Demircan:OL:2014}
\bibinfo{author}{\bibfnamefont{A.}~\bibnamefont{Demircan}},
  \bibinfo{author}{\bibfnamefont{S.}~\bibnamefont{Amiranashvili}},
  \bibinfo{author}{\bibfnamefont{C.}~\bibnamefont{Br\'ee}},
  \bibinfo{author}{\bibfnamefont{U.}~\bibnamefont{Morgner}}, \bibnamefont{and}
  \bibinfo{author}{\bibfnamefont{G.}~\bibnamefont{Steinmeyer}},
  \bibinfo{journal}{Opt. Lett.} \textbf{\bibinfo{volume}{39}},
  \bibinfo{pages}{2735} (\bibinfo{year}{2014}{\natexlab{a}}).

\bibitem[{\citenamefont{Driben et~al.}(2010)\citenamefont{Driben, Mitschke, and
  Zhavoronkov}}]{Driben:OE:2010}
\bibinfo{author}{\bibfnamefont{R.}~\bibnamefont{Driben}},
  \bibinfo{author}{\bibfnamefont{F.}~\bibnamefont{Mitschke}}, \bibnamefont{and}
  \bibinfo{author}{\bibfnamefont{N.}~\bibnamefont{Zhavoronkov}},
  \bibinfo{journal}{Opt. Express} \textbf{\bibinfo{volume}{18}},
  \bibinfo{pages}{25993} (\bibinfo{year}{2010}).

\bibitem[{\citenamefont{Demircan
  et~al.}(2014{\natexlab{b}})\citenamefont{Demircan, Amiranashvili, Br\'{e}e,
  Mahnke, Mitschke, and Steinmeyer}}]{Demircan:APB:2014}
\bibinfo{author}{\bibfnamefont{A.}~\bibnamefont{Demircan}},
  \bibinfo{author}{\bibfnamefont{S.}~\bibnamefont{Amiranashvili}},
  \bibinfo{author}{\bibfnamefont{C.}~\bibnamefont{Br\'{e}e}},
  \bibinfo{author}{\bibfnamefont{C.}~\bibnamefont{Mahnke}},
  \bibinfo{author}{\bibfnamefont{F.}~\bibnamefont{Mitschke}}, \bibnamefont{and}
  \bibinfo{author}{\bibfnamefont{G.}~\bibnamefont{Steinmeyer}},
  \bibinfo{journal}{Appl. Phys. B} \textbf{\bibinfo{volume}{115}},
  \bibinfo{pages}{343–354} (\bibinfo{year}{2014}{\natexlab{b}}).

\bibitem[{\citenamefont{Melchert
  et~al.}(2019{\natexlab{a}})\citenamefont{Melchert, Willms, Bose, Yulin, Roth,
  Mitschke, Morgner, Babushkin, and Demircan}}]{Melchert:PRL:2019}
\bibinfo{author}{\bibfnamefont{O.}~\bibnamefont{Melchert}},
  \bibinfo{author}{\bibfnamefont{S.}~\bibnamefont{Willms}},
  \bibinfo{author}{\bibfnamefont{S.}~\bibnamefont{Bose}},
  \bibinfo{author}{\bibfnamefont{A.}~\bibnamefont{Yulin}},
  \bibinfo{author}{\bibfnamefont{B.}~\bibnamefont{Roth}},
  \bibinfo{author}{\bibfnamefont{F.}~\bibnamefont{Mitschke}},
  \bibinfo{author}{\bibfnamefont{U.}~\bibnamefont{Morgner}},
  \bibinfo{author}{\bibfnamefont{I.}~\bibnamefont{Babushkin}},
  \bibnamefont{and} \bibinfo{author}{\bibfnamefont{A.}~\bibnamefont{Demircan}},
  \bibinfo{journal}{Phys. Rev. Lett.} \textbf{\bibinfo{volume}{123}},
  \bibinfo{pages}{243905} (\bibinfo{year}{2019}{\natexlab{a}}).

\bibitem[{\citenamefont{Melchert and Demircan}(2021)}]{Melchert:OL:2021}
\bibinfo{author}{\bibfnamefont{O.}~\bibnamefont{Melchert}} \bibnamefont{and}
  \bibinfo{author}{\bibfnamefont{A.}~\bibnamefont{Demircan}},
  \bibinfo{journal}{Opt. Lett.} \textbf{\bibinfo{volume}{46}},
  \bibinfo{pages}{5603} (\bibinfo{year}{2021}).

\bibitem[{\citenamefont{Melchert et~al.}(2021)\citenamefont{Melchert, Willms,
  Morgner, Babushkin, and Demircan}}]{Melchert:SR:2021}
\bibinfo{author}{\bibfnamefont{O.}~\bibnamefont{Melchert}},
  \bibinfo{author}{\bibfnamefont{S.}~\bibnamefont{Willms}},
  \bibinfo{author}{\bibfnamefont{U.}~\bibnamefont{Morgner}},
  \bibinfo{author}{\bibfnamefont{I.}~\bibnamefont{Babushkin}},
  \bibnamefont{and} \bibinfo{author}{\bibfnamefont{A.}~\bibnamefont{Demircan}},
  \bibinfo{journal}{Scientific Reports} \textbf{\bibinfo{volume}{11}},
  \bibinfo{pages}{11190} (\bibinfo{year}{2021}).

\bibitem[{\citenamefont{Willms et~al.}(2022)\citenamefont{Willms, Melchert,
  Bose, Yulin, Oreshnikov, Morgner, Babushkin, and Demircan}}]{Willms:PRA:2022}
\bibinfo{author}{\bibfnamefont{S.}~\bibnamefont{Willms}},
  \bibinfo{author}{\bibfnamefont{O.}~\bibnamefont{Melchert}},
  \bibinfo{author}{\bibfnamefont{S.}~\bibnamefont{Bose}},
  \bibinfo{author}{\bibfnamefont{A.}~\bibnamefont{Yulin}},
  \bibinfo{author}{\bibfnamefont{I.}~\bibnamefont{Oreshnikov}},
  \bibinfo{author}{\bibfnamefont{U.}~\bibnamefont{Morgner}},
  \bibinfo{author}{\bibfnamefont{I.}~\bibnamefont{Babushkin}},
  \bibnamefont{and} \bibinfo{author}{\bibfnamefont{A.}~\bibnamefont{Demircan}},
  \bibinfo{journal}{Phys. Rev. A} \textbf{\bibinfo{volume}{105}},
  \bibinfo{pages}{053525} (\bibinfo{year}{2022}).

\bibitem[{\citenamefont{Oreshnikov et~al.}(2022)\citenamefont{Oreshnikov,
  Melchert, Willms, Bose, Babushkin, Demircan, Morgner, and
  Yulin}}]{Oreshnikov:arXiv:2022}
\bibinfo{author}{\bibfnamefont{I.}~\bibnamefont{Oreshnikov}},
  \bibinfo{author}{\bibfnamefont{O.}~\bibnamefont{Melchert}},
  \bibinfo{author}{\bibfnamefont{S.}~\bibnamefont{Willms}},
  \bibinfo{author}{\bibfnamefont{S.}~\bibnamefont{Bose}},
  \bibinfo{author}{\bibfnamefont{I.}~\bibnamefont{Babushkin}},
  \bibinfo{author}{\bibfnamefont{A.}~\bibnamefont{Demircan}},
  \bibinfo{author}{\bibfnamefont{U.}~\bibnamefont{Morgner}}, \bibnamefont{and}
  \bibinfo{author}{\bibfnamefont{A.}~\bibnamefont{Yulin}},
  \emph{\bibinfo{title}{{Cherenkov radiation and scattering of external
  dispersive waves by two-color solitons}}} (\bibinfo{year}{2022}),
  \bibinfo{note}{arXiv:2207.03541},
  \urlprefix\url{{https://doi.org/10.48550/arXiv.2207.03541}}.

\bibitem[{\citenamefont{Tam et~al.}(2019)\citenamefont{Tam, Alexander,
  Blanco-Redondo, and de~Sterke}}]{Tam:OL:2019}
\bibinfo{author}{\bibfnamefont{K.~K.~K.} \bibnamefont{Tam}},
  \bibinfo{author}{\bibfnamefont{T.~J.} \bibnamefont{Alexander}},
  \bibinfo{author}{\bibfnamefont{A.}~\bibnamefont{Blanco-Redondo}},
  \bibnamefont{and} \bibinfo{author}{\bibfnamefont{C.~M.}
  \bibnamefont{de~Sterke}}, \bibinfo{journal}{Opt. Lett.}
  \textbf{\bibinfo{volume}{44}}, \bibinfo{pages}{3306} (\bibinfo{year}{2019}).

\bibitem[{\citenamefont{Lourdesamy et~al.}(2021)\citenamefont{Lourdesamy,
  Runge, Alexander, Hudson, Blanco-Redondo, and
  de~Sterke}}]{Lourdesamy:NP:2021}
\bibinfo{author}{\bibfnamefont{J.~P.} \bibnamefont{Lourdesamy}},
  \bibinfo{author}{\bibfnamefont{A.~F.~J.} \bibnamefont{Runge}},
  \bibinfo{author}{\bibfnamefont{T.~J.} \bibnamefont{Alexander}},
  \bibinfo{author}{\bibfnamefont{D.~D.} \bibnamefont{Hudson}},
  \bibinfo{author}{\bibfnamefont{A.}~\bibnamefont{Blanco-Redondo}},
  \bibnamefont{and} \bibinfo{author}{\bibfnamefont{C.~M.}
  \bibnamefont{de~Sterke}}, \bibinfo{journal}{Nat. Phys.}
  \textbf{\bibinfo{volume}{18}}, \bibinfo{pages}{59} (\bibinfo{year}{2021}).

\bibitem[{\citenamefont{Mao et~al.}(2021)\citenamefont{Mao, Wang, Zhang, Zeng,
  Du, He, Sun, and Zhao}}]{Mao:NC:2021}
\bibinfo{author}{\bibfnamefont{D.}~\bibnamefont{Mao}},
  \bibinfo{author}{\bibfnamefont{H.}~\bibnamefont{Wang}},
  \bibinfo{author}{\bibfnamefont{H.}~\bibnamefont{Zhang}},
  \bibinfo{author}{\bibfnamefont{C.}~\bibnamefont{Zeng}},
  \bibinfo{author}{\bibfnamefont{Y.}~\bibnamefont{Du}},
  \bibinfo{author}{\bibfnamefont{Z.}~\bibnamefont{He}},
  \bibinfo{author}{\bibfnamefont{Z.}~\bibnamefont{Sun}}, \bibnamefont{and}
  \bibinfo{author}{\bibfnamefont{J.}~\bibnamefont{Zhao}},
  \bibinfo{journal}{Nat. Commun.} \textbf{\bibinfo{volume}{12}},
  \bibinfo{pages}{6712} (\bibinfo{year}{2021}).

\bibitem[{\citenamefont{Driben et~al.}(2009)\citenamefont{Driben, Husakou, and
  Herrmann}}]{Driben:OE:2009}
\bibinfo{author}{\bibfnamefont{R.}~\bibnamefont{Driben}},
  \bibinfo{author}{\bibfnamefont{A.}~\bibnamefont{Husakou}}, \bibnamefont{and}
  \bibinfo{author}{\bibfnamefont{J.}~\bibnamefont{Herrmann}},
  \bibinfo{journal}{Opt. Express} \textbf{\bibinfo{volume}{17}},
  \bibinfo{pages}{17989} (\bibinfo{year}{2009}).

\bibitem[{\citenamefont{Bose et~al.}(2016{\natexlab{a}})\citenamefont{Bose,
  Sahoo, Chattopadhyay, Roy, Bhadra, and Agrawal}}]{Bose:PRA:2016}
\bibinfo{author}{\bibfnamefont{S.}~\bibnamefont{Bose}},
  \bibinfo{author}{\bibfnamefont{A.}~\bibnamefont{Sahoo}},
  \bibinfo{author}{\bibfnamefont{R.}~\bibnamefont{Chattopadhyay}},
  \bibinfo{author}{\bibfnamefont{S.}~\bibnamefont{Roy}},
  \bibinfo{author}{\bibfnamefont{S.~K.} \bibnamefont{Bhadra}},
  \bibnamefont{and} \bibinfo{author}{\bibfnamefont{G.~P.}
  \bibnamefont{Agrawal}}, \bibinfo{journal}{Phys. Rev. A}
  \textbf{\bibinfo{volume}{94}}, \bibinfo{pages}{043835}
  (\bibinfo{year}{2016}{\natexlab{a}}).

\bibitem[{\citenamefont{Bose et~al.}(2016{\natexlab{b}})\citenamefont{Bose,
  Chattopadhyay, Roy, and Bhadra}}]{Bose:JOSAB:2016}
\bibinfo{author}{\bibfnamefont{S.}~\bibnamefont{Bose}},
  \bibinfo{author}{\bibfnamefont{R.}~\bibnamefont{Chattopadhyay}},
  \bibinfo{author}{\bibfnamefont{S.}~\bibnamefont{Roy}}, \bibnamefont{and}
  \bibinfo{author}{\bibfnamefont{S.~K.} \bibnamefont{Bhadra}},
  \bibinfo{journal}{J. Opt. Soc. Am. B} \textbf{\bibinfo{volume}{33}},
  \bibinfo{pages}{1014} (\bibinfo{year}{2016}{\natexlab{b}}).

\bibitem[{\citenamefont{Arteaga-Sierra
  et~al.}(2018)\citenamefont{Arteaga-Sierra, Antikainen, and
  Agrawal}}]{Arteaga:PRA:2018}
\bibinfo{author}{\bibfnamefont{F.~R.} \bibnamefont{Arteaga-Sierra}},
  \bibinfo{author}{\bibfnamefont{A.}~\bibnamefont{Antikainen}},
  \bibnamefont{and} \bibinfo{author}{\bibfnamefont{G.~P.}
  \bibnamefont{Agrawal}}, \bibinfo{journal}{Phys. Rev. A}
  \textbf{\bibinfo{volume}{98}}, \bibinfo{pages}{013830}
  (\bibinfo{year}{2018}).

\bibitem[{\citenamefont{Linale et~al.}(2020)\citenamefont{Linale, Bonetti,
  S\'{a}nchez, Hernandez, Fierens, and Grosz}}]{Linale:OL:2020}
\bibinfo{author}{\bibfnamefont{N.}~\bibnamefont{Linale}},
  \bibinfo{author}{\bibfnamefont{J.}~\bibnamefont{Bonetti}},
  \bibinfo{author}{\bibfnamefont{A.~D.} \bibnamefont{S\'{a}nchez}},
  \bibinfo{author}{\bibfnamefont{S.}~\bibnamefont{Hernandez}},
  \bibinfo{author}{\bibfnamefont{P.~I.} \bibnamefont{Fierens}},
  \bibnamefont{and} \bibinfo{author}{\bibfnamefont{D.~F.} \bibnamefont{Grosz}},
  \bibinfo{journal}{Opt. Lett.} \textbf{\bibinfo{volume}{45}},
  \bibinfo{pages}{2498} (\bibinfo{year}{2020}).

\bibitem[{\citenamefont{Hernandez et~al.}(2022)\citenamefont{Hernandez,
  Sparapani, Linale, Bonetti, Grosz, and Fierens}}]{Hernandez:WRCM:2022}
\bibinfo{author}{\bibfnamefont{S.~M.} \bibnamefont{Hernandez}},
  \bibinfo{author}{\bibfnamefont{A.}~\bibnamefont{Sparapani}},
  \bibinfo{author}{\bibfnamefont{N.}~\bibnamefont{Linale}},
  \bibinfo{author}{\bibfnamefont{J.}~\bibnamefont{Bonetti}},
  \bibinfo{author}{\bibfnamefont{D.~F.} \bibnamefont{Grosz}}, \bibnamefont{and}
  \bibinfo{author}{\bibfnamefont{P.~I.} \bibnamefont{Fierens}},
  \bibinfo{journal}{Waves in Random and Complex Media} pp.
  \bibinfo{pages}{1--15} (\bibinfo{year}{2022}).

\bibitem[{\citenamefont{Junnarkar and Uesugi}(2000)}]{Junnarkar:OC:2000}
\bibinfo{author}{\bibfnamefont{M.~R.} \bibnamefont{Junnarkar}}
  \bibnamefont{and} \bibinfo{author}{\bibfnamefont{N.}~\bibnamefont{Uesugi}},
  \bibinfo{journal}{Opt. Commun.} \textbf{\bibinfo{volume}{175}},
  \bibinfo{pages}{447} (\bibinfo{year}{2000}).

\bibitem[{\citenamefont{Agrawal}(2019)}]{Agrawal:BOOK:2019}
\bibinfo{author}{\bibfnamefont{G.~P.} \bibnamefont{Agrawal}},
  \emph{\bibinfo{title}{Nonlinear Fiber Optics}} (\bibinfo{publisher}{Academic
  Press}, \bibinfo{year}{2019}).

\bibitem[{\citenamefont{Zhao et~al.}(2022)\citenamefont{Zhao, Guo, and
  Zeng}}]{Zhao:PRA:2022}
\bibinfo{author}{\bibfnamefont{S.}~\bibnamefont{Zhao}},
  \bibinfo{author}{\bibfnamefont{R.}~\bibnamefont{Guo}}, \bibnamefont{and}
  \bibinfo{author}{\bibfnamefont{Y.}~\bibnamefont{Zeng}},
  \bibinfo{journal}{Phys. Rev. A} \textbf{\bibinfo{volume}{106}},
  \bibinfo{pages}{033516} (\bibinfo{year}{2022}).

\bibitem[{\citenamefont{Blow and Wood}(1989)}]{Blow:JQE:1989}
\bibinfo{author}{\bibfnamefont{K.~J.} \bibnamefont{Blow}} \bibnamefont{and}
  \bibinfo{author}{\bibfnamefont{D.}~\bibnamefont{Wood}},
  \bibinfo{journal}{IEEE J. Quantum Electron.} \textbf{\bibinfo{volume}{25}},
  \bibinfo{pages}{2665} (\bibinfo{year}{1989}).

\bibitem[{\citenamefont{Heidt}(2009)}]{Heidt:JLT:2009}
\bibinfo{author}{\bibfnamefont{A.~M.} \bibnamefont{Heidt}},
  \bibinfo{journal}{IEEE J. Lightwave Tech.} \textbf{\bibinfo{volume}{27}},
  \bibinfo{pages}{3984} (\bibinfo{year}{2009}).

\bibitem[{\citenamefont{Melchert and Demircan}(2022)}]{Melchert:CPC:2022}
\bibinfo{author}{\bibfnamefont{O.}~\bibnamefont{Melchert}} \bibnamefont{and}
  \bibinfo{author}{\bibfnamefont{A.}~\bibnamefont{Demircan}},
  \bibinfo{journal}{Computer Physics Communications}
  \textbf{\bibinfo{volume}{273}}, \bibinfo{pages}{108257}
  (\bibinfo{year}{2022}).

\bibitem[{\citenamefont{Melchert
  et~al.}(2019{\natexlab{b}})\citenamefont{Melchert, Roth, Morgner, and
  Demircan}}]{Melchert:SFX:2019}
\bibinfo{author}{\bibfnamefont{O.}~\bibnamefont{Melchert}},
  \bibinfo{author}{\bibfnamefont{B.}~\bibnamefont{Roth}},
  \bibinfo{author}{\bibfnamefont{U.}~\bibnamefont{Morgner}}, \bibnamefont{and}
  \bibinfo{author}{\bibfnamefont{A.}~\bibnamefont{Demircan}},
  \bibinfo{journal}{SoftwareX} \textbf{\bibinfo{volume}{10}},
  \bibinfo{pages}{100275} (\bibinfo{year}{2019}{\natexlab{b}}).

\bibitem[{\citenamefont{Landau and Lifshitz}(1981)}]{Landau:BOOK:1981}
\bibinfo{author}{\bibfnamefont{L.~D.} \bibnamefont{Landau}} \bibnamefont{and}
  \bibinfo{author}{\bibfnamefont{L.~M.} \bibnamefont{Lifshitz}},
  \emph{\bibinfo{title}{Quantum Mechanics Non-Relativistic Theory, Third
  Edition: Volume 3}} (\bibinfo{publisher}{Elsevier Science, Oxford},
  \bibinfo{year}{1981}).

\bibitem[{\citenamefont{Lekner}(2007)}]{Lekner:AJP:2007}
\bibinfo{author}{\bibfnamefont{J.}~\bibnamefont{Lekner}}, \bibinfo{journal}{Am.
  J. Phys.} \textbf{\bibinfo{volume}{75}}, \bibinfo{pages}{1151}
  (\bibinfo{year}{2007}).

\bibitem[{\citenamefont{Haus and Ippen}(2001)}]{Haus:OL:2001}
\bibinfo{author}{\bibfnamefont{H.~A.} \bibnamefont{Haus}} \bibnamefont{and}
  \bibinfo{author}{\bibfnamefont{E.~P.} \bibnamefont{Ippen}},
  \bibinfo{journal}{Opt. Lett.} \textbf{\bibinfo{volume}{26}},
  \bibinfo{pages}{1654} (\bibinfo{year}{2001}).

\bibitem[{\citenamefont{Haelterman et~al.}(1993)\citenamefont{Haelterman,
  Sheppard, and Snyder}}]{Haelterman:OL:1993}
\bibinfo{author}{\bibfnamefont{M.}~\bibnamefont{Haelterman}},
  \bibinfo{author}{\bibfnamefont{A.}~\bibnamefont{Sheppard}}, \bibnamefont{and}
  \bibinfo{author}{\bibfnamefont{A.}~\bibnamefont{Snyder}},
  \bibinfo{journal}{Opt. Lett.} \textbf{\bibinfo{volume}{18}},
  \bibinfo{pages}{1406} (\bibinfo{year}{1993}).

\bibitem[{\citenamefont{Silberberg and Barad}(1995)}]{Silberberg:OL:1995}
\bibinfo{author}{\bibfnamefont{Y.}~\bibnamefont{Silberberg}} \bibnamefont{and}
  \bibinfo{author}{\bibfnamefont{Y.}~\bibnamefont{Barad}},
  \bibinfo{journal}{Opt. Lett.} \textbf{\bibinfo{volume}{20}},
  \bibinfo{pages}{246} (\bibinfo{year}{1995}).

\bibitem[{\citenamefont{Afanasyev et~al.}(1989)\citenamefont{Afanasyev,
  Kivshar, Konotop, and Serkin}}]{Afanasjev:OL:1989}
\bibinfo{author}{\bibfnamefont{V.~V.} \bibnamefont{Afanasyev}},
  \bibinfo{author}{\bibfnamefont{Y.~S.} \bibnamefont{Kivshar}},
  \bibinfo{author}{\bibfnamefont{V.~V.} \bibnamefont{Konotop}},
  \bibnamefont{and} \bibinfo{author}{\bibfnamefont{V.~N.}
  \bibnamefont{Serkin}}, \bibinfo{journal}{Opt. Lett.}
  \textbf{\bibinfo{volume}{14}}, \bibinfo{pages}{805} (\bibinfo{year}{1989}).

\bibitem[{\citenamefont{Pelinovsky and Kivshar}(2000)}]{Pelinovsky:PRE:2000}
\bibinfo{author}{\bibfnamefont{D.}~\bibnamefont{Pelinovsky}} \bibnamefont{and}
  \bibinfo{author}{\bibfnamefont{Y.}~\bibnamefont{Kivshar}},
  \bibinfo{journal}{Phys. Rev. E} \textbf{\bibinfo{volume}{62}},
  \bibinfo{pages}{8668} (\bibinfo{year}{2000}).

\bibitem[{\citenamefont{Haelterman and Sheppard}(1994)}]{Haelterman:PRE:1994}
\bibinfo{author}{\bibfnamefont{M.}~\bibnamefont{Haelterman}} \bibnamefont{and}
  \bibinfo{author}{\bibfnamefont{A.}~\bibnamefont{Sheppard}},
  \bibinfo{journal}{Phys. Rev. E} \textbf{\bibinfo{volume}{149}},
  \bibinfo{pages}{3376} (\bibinfo{year}{1994}).

\bibitem[{\citenamefont{Mitchell et~al.}(1997)\citenamefont{Mitchell, Segev,
  Coskun, and Christodulides}}]{Mitchell:PRL:1997}
\bibinfo{author}{\bibfnamefont{M.}~\bibnamefont{Mitchell}},
  \bibinfo{author}{\bibfnamefont{M.}~\bibnamefont{Segev}},
  \bibinfo{author}{\bibfnamefont{T.}~\bibnamefont{Coskun}}, \bibnamefont{and}
  \bibinfo{author}{\bibfnamefont{D.}~\bibnamefont{Christodulides}},
  \bibinfo{journal}{Phys. Rev. Lett.} \textbf{\bibinfo{volume}{79}},
  \bibinfo{pages}{4990} (\bibinfo{year}{1997}).

\bibitem[{\citenamefont{Ablowitz and Musslimani}(2005)}]{Ablowitz:OL:2005}
\bibinfo{author}{\bibfnamefont{M.}~\bibnamefont{Ablowitz}} \bibnamefont{and}
  \bibinfo{author}{\bibfnamefont{Z.}~\bibnamefont{Musslimani}},
  \bibinfo{journal}{Opt. Lett.} \textbf{\bibinfo{volume}{30}},
  \bibinfo{pages}{2140} (\bibinfo{year}{2005}).

\bibitem[{\citenamefont{Lakoba and Yang}(2007)}]{Lakoba:JCP:2007}
\bibinfo{author}{\bibfnamefont{T.}~\bibnamefont{Lakoba}} \bibnamefont{and}
  \bibinfo{author}{\bibfnamefont{J.}~\bibnamefont{Yang}}, \bibinfo{journal}{J.
  Comp. Phys.} \textbf{\bibinfo{volume}{226}}, \bibinfo{pages}{1668}
  (\bibinfo{year}{2007}).

\bibitem[{\citenamefont{Lakoba}(2009)}]{Lakoba:PD:2009}
\bibinfo{author}{\bibfnamefont{T.}~\bibnamefont{Lakoba}},
  \bibinfo{journal}{Physica D} \textbf{\bibinfo{volume}{238}},
  \bibinfo{pages}{2308} (\bibinfo{year}{2009}).

\bibitem[{\citenamefont{Yang}(2009)}]{Yang:JCP:2009}
\bibinfo{author}{\bibfnamefont{J.}~\bibnamefont{Yang}}, \bibinfo{journal}{J.
  Comp. Phys.} \textbf{\bibinfo{volume}{228}}, \bibinfo{pages}{7007}
  (\bibinfo{year}{2009}).

\bibitem[{\citenamefont{Dror and Malomed}(2016)}]{Dror:JO:2016}
\bibinfo{author}{\bibfnamefont{N.}~\bibnamefont{Dror}} \bibnamefont{and}
  \bibinfo{author}{\bibfnamefont{B.}~\bibnamefont{Malomed}},
  \bibinfo{journal}{J. Opt.} \textbf{\bibinfo{volume}{18}},
  \bibinfo{pages}{014003} (\bibinfo{year}{2016}).

\bibitem[{\citenamefont{Kierzenka and Shampine}(2001)}]{Kierzenka:ACM:2001}
\bibinfo{author}{\bibfnamefont{J.}~\bibnamefont{Kierzenka}} \bibnamefont{and}
  \bibinfo{author}{\bibfnamefont{L.}~\bibnamefont{Shampine}},
  \bibinfo{journal}{ACM Trans. Math. Softw.} \textbf{\bibinfo{volume}{27}},
  \bibinfo{pages}{300} (\bibinfo{year}{2001}).

\end{thebibliography}

\end{document}